# Contact-Based Architecture for Resource Discovery (*CARD*) in Large Scale MANets

Saurabh Garg, Priyatham Pamu, Nitin Nahata, Ahmed Helmy
University of Southern California
{sgarg, pamu, nnahata, helmy}@usc.edu

***Abstract* - In this paper we propose a novel architecture, *CARD*, for resource discovery in large scale Mobile Ad hoc Networks (MANets) which, may scale up to thousands of nodes and may span wide geographical regions. Unlike previously proposed schemes, our architecture avoids expensive mechanisms such as global flooding as well as complex coordination between nodes to form a hierarchy. *CARD* is also independent of any external source of information such as GPS. In our architecture nodes within a limited number of hops from each node form the *neighborhood* of that node. Resources within the neighborhood can be readily accessed with the help of a proactive scheme within the neighborhood. For accessing resources beyond the neighborhood, each node also maintains a few distant nodes called *contacts*. Contacts help in creating a *small world* in the network and provide an efficient way to query for resources beyond the neighborhood. As the number of contacts of a node increases, the network view (reachability) of the node increases. Paths to contacts are validated periodically to adapt to mobility. We present mechanisms for contact selection and maintenance that attempt to increase reachability while minimizing overhead. Our simulation results show a clear trade-off between increase in reachability on one hand, and contact selection and maintenance overhead on the other. Our results suggest that *CARD* can be configured to provide a desirable reachability distribution for different network sizes. Comparisons with other schemes for resource discovery, such as flooding and *bordercasting*, show our architecture to be much more efficient and scalable.**

## I. INTRODUCTION

Ad hoc networks are wireless networks composed of mobile devices with limited power and transmission range. These networks are rapidly deployable as they neither require a wired infrastructure nor centralized control. Because of the lack of fixed infrastructure, each node also acts as a relay to provide communication throughout the network. Applications of ad hoc networks include coordination between various units (e.g., in a battlefield), search and rescue missions, rapidly deployable networks, and vehicular networks, among others.

Lack of a fixed infrastructure, makes resource discovery in ad hoc networks a challenging problem. Unlike wired networks, mobility induces frequent route changes. Traditional protocols proposed for resource discovery in such environments either involve global flooding or are based on complex hierarchy formation. While flooding is inefficient and therefore does not scale well, hierarchy formation involves complex coordination between nodes and therefore may suffer significant performance degradation due to frequent, mobility induced, changes in network connectivity.

To overcome these limitations we propose a new architecture based on the concept of *small worlds* [10] [11]. In our architecture we adopt a hybrid approach in which a node uses periodic updates to reach its *neighborhood* within a limited number of hops, *R*, and reactive querying beyond the neighborhood via *contacts*. *Contacts* act as *short cuts* that attempt to transform the network into a small world by reducing the degrees of separation. They help in providing a view of the network beyond the neighborhood during resource discovery. Each node maintains state for a few contacts beyond its neighborhood. Contacts are polled periodically to validate their presence and routes. For discovering resources efficiently, queries are sent to the contacts that leverage the knowledge of their neighborhood. As the number of contacts increases, the network view (reachability) increases. However, at the same time the overhead involved in contact maintenance also increases. Our simulation results show this trade-off.

Our architecture has been designed to satisfy requirements for an efficient resource discovery scheme. Applications like sensor networks may comprise of thousands of nodes. Therefore the resource discovery mechanism should be scalable. Nodes in an ad hoc network comprise of portable devices with limited battery power. Therefore to save power the resource discovery mechanism should be efficient in terms of messages transmitted. Simulation based comparisons with flooding and bordercasting [8][9] show our architecture to be more efficient. Simulation results also show that our protocol is scalable and can be configured to provide good performance for various network sizes.

The rest of this document is organized as follows. Section II discusses related work. Section III describes our design requirements and provides an overview of our proposed architecture, *CARD*, and introduces the contact selection and maintenance algorithms. Section IV presents thorough analysis of *CARD*, in terms of reachability and overhead, and compares it to other schemes such as flooding and bordercasting. We conclude in Section V.

## II. RELATED WORK

Due to lack of an existing infrastructure in ad hoc networks, resource discovery (which includes route discovery) is a challenging problem. Most of the protocols proposed so far can be broadly classified as: (1) Proactive (table driven), reactive (on - demand) and hybrid, or (2) flat and hierarchical.

Proactive schemes such as DSDV [1], WRP [3] and GSR [2] involve periodic updates to be flooded throughout the network. This is a resource consuming process especially for large-scale networks that may scale up to thousands of nodes. Reactive schemes such as AODV [5] and DSR [4] try to reduce the overhead due to periodic updates by maintaining state only for the active resources. In these schemes a search is initiated for any new resource required. However, the search procedure generally involves flooding through the entire network. This is again inefficient and also leads to increased response delays for resource discovery. The above architectures are flat (i.e., non-hierarchical), in which each node has an equal responsibility of relaying traffic from other nodes. Such schemes are expensive in terms of resources and do not scale well.

Hybrid schemes such as ZRP [9] try to combine the benefits of both the proactive and reactive schemes. ZRP limits the overhead of periodic updates to a limited number of hops (zone). Resources beyond the zone are searched in a reactive manner by sending queries through the nodes at the edge of the zone (bordercasting). The zone concept is similar to the neighborhood concept in our study. However, instead of bordercasting we use contacts. In our study, we will show that the contact-based approach is more efficient than bordercasting.

Hierarchical schemes, on the other hand, such as CGSR [6] and [15], involve election of a cluster-head, which has greater responsibilities than other nodes. The cluster-head is responsible for routing traffic in and out of the cluster. Cluster-based hierarchies rely on complex coordination and thus are susceptible to major re-configuration due to mobility. This could lead to serious performance degradation. Also, a cluster head may be a single point of failure and a potential bottleneck. In our architecture there is very simple coordination between various nodes. This enables our architecture to adapt gracefully to network dynamics. Each node has its own view of the network and hence does not require major re-configuration with mobility.

Architectures such as GLS [7] -that require knowledge by all the nodes of a global grid mapping the whole network- have also been proposed. GLS requires knowledge of geographic location (obtained from an external source such as GPS). Our architecture is independent of any external information such as GPS.

In [12] a framework was proposed for multicast in large-scale ad hoc networks that introduced the concept of contacts. However, only high level concepts were proposed and no detailed algorithms or evaluations were presented for contact selection and maintenance. Our work here fits nicely in that framework. Also, in [13] the relationship between small worlds and wireless networks was established. In this paper, we build upon that relationship with small worlds. Furthermore, a mobility-assisted scheme was proposed for contact selection in that study. Although such scheme could be very efficient, it depends heavily on mobility, and may not be suitable for static sensor networks. The algorithms presented in this paper may complement those proposed in the above work, and may be integrated with them.

## III. *CARD* ARCHITECTURAL OVERVIEW

### *A. Design Requirements*

The requirements for a resource discovery mechanism for large-scale Ad hoc networks that motivate the design of our architecture include:

*(a) Scalability*: Applications of large-scale ad hoc networks involve military and sensor network environments that may span wide geographical areas and may comprise of thousands of nodes. Therefore the resource discovery mechanism should be scalable in terms of control overhead with increase in network size.

*(b) Efficiency*: Nodes in an ad hoc network comprise of portable devices with limited battery power. Therefore, resource discovery mechanisms should be efficient.

*(c) Robustness*: The mechanism should be robust to handle frequent link failures due to mobility.

*(d) Decentralized operation*: For the network to be rapidly deployable, it should not require any centralized control.

*(e) No location (geographic) information*: Rapid deployability and self-configurability require that the

network mechanism should be independent of any information that requires an external source.

*B. Definitions*

• *Neighborhood:* All nodes within a particular number of hops (*R*) from the source node. *R* is the radius of the neighborhood.
• *Edge nodes:* All nodes at a distance of *R* hops from the source node.
• *Maximum contact distance (r):* This is the maximum distance (in hops) from the source within which a contact is selected.
• *Overlap:* Overlap between nodes represents number of common nodes between their neighborhoods.
• *Number of Contacts (NoC): NoC* specifies the value of the maximum number of contacts to be searched for each source node. The actual number of contacts chosen is usually less than this value. This is due to the fact that for a particular value of *R* and *r*, there is only a limited region available for choosing contacts. Once this region has been covered by neighborhoods of the chosen contacts, choosing more contacts in the same region is not possible, as their neighborhoods would overlap with the neighborhoods of the already chosen contacts. This is according to our policy to minimize overlap.
• *Depth of search (D): D* specifies the levels of contacts (i.e., contacts of contacts) queried by a source.
• *Reachability:* The number of nodes that can be reached by a source node is termed as the reachability of the source node. This includes the nodes within the neighborhood that can be reached directly and the nodes that lie in the contacts neighborhood and are therefore reachable through the contact(s), etc.
• *Overhead:* Overhead is defined in terms of the control messages generated by a mechanism. For example, contact selection, maintenance or query. Procedures for these mechanisms are described in the next sections.

*C. Mechanism Description*

Our mechanism employs a hybrid of proactive and reactive approaches for resource discovery. All nodes within *R* hops from a node form the node's neighborhood. Each node proactively (using a protocol such as DSDV[1]) maintains state for all the nodes in its neighborhood. Therefore a node has complete knowledge of all the nodes (resources) within its neighborhood. Apart from the neighborhood knowledge, each node also maintains state for a few nodes which lie outside the neighborhood. These nodes serve as contacts for accessing resources beyond the neighborhood. Contacts are selected and maintained using the mechanisms described below.

***1. Contact Selection Procedure***

*(1)* Each node (source node ***s***) sends a Contact Selection Query (*CSQ*) through each of its edge node, one at a time.
*(2)* An edge node receiving a contact selection query forwards *CSQ* to a randomly chosen neighbor (*X*).
*(3)* A node receiving a *CSQ* decides whether or not to be a contact for the particular source. This decision is made using either a probabilistic method (*PM*) or edge method (*EM*). These methods are described later in this section.
*(4)* After using either procedure *PM* or *EM* for deciding whether to be a contact, if the node receiving the *CSQ* does not choose to be the contact, it forwards the query to one of its randomly chosen neighbor (excluding the one from which *CSQ* was received).
*(5)* The *CSQ* traverses in a depth-first manner until a contact is chosen or it reaches a node at a distance of *r* hops from the source. If a contact is still not chosen (due to overlap), the query backtracks to the previous node, which forwards it to another randomly chosen neighbor.
*(6)* When a contact is selected, the path to the contact is returned and stored at the source node.

***2. Contact Selection Methods***

We discuss and compare two different methods or protocols for contact selection: (a) the probabilistic method (*PM*), and (b) the edge method (*EM*).

*(a) Probabilistic Method (PM):*
Contacts help to increase a node's view (or reachability) of the network beyond its own neighborhood. To achieve maximum increase in reachability, the neighborhoods of any source and its contacts should be disjoint/separated, i.e., there should be no overlap between the neighborhood of the source and the neighborhood of any of its contacts. The neighborhoods of different contacts of the same source should also be non-overlapping, to ensure maximum increase in reachability. To achieve this, the *CSQ* contains the following information: (*i*) ID of the source node, (*ii*) a list of already chosen contacts of the source node (*Contact_List*; typically small of ~5 IDs*)*, and (*iii)* the hop count, *d*.
This information is used as follows. When a node *X* receives a *CSQ*, it first checks if the source lies within its neighborhood. This check is easily performed since each node has complete knowledge of its neighborhood. So a node knows the IDs of all the other nodes in its neighborhood. *X* also checks if its neighborhood contains any of the node IDs contained in the *Contact_List.*

If neither the source nor any of its already selected contacts lie in the neighborhood of *X*, *X* probabilistically chooses itself as the contact. This probability (*P*) of choosing to be a contact is defined as follows:

$$P = (d - R)/(r - R) \quad \text{-- (1)}$$

Here d is the distance of *X* from the source node. *d* is included in the *CSQ* as hop count. From the above equation, when $d = R$, $P = 0$. When $d = r$, $P = 1$. This aims to select the contacts between *R* and *r* hops from the source. This has been formulated to provide maximum increase in reachablility with the addition of each new contact. However there can still be a case where equation (*1*) does not provide the maximum benefit of adding a contact. This case is shown in the fig.1 where ***s*** is the source node and ***c*** is the contact.

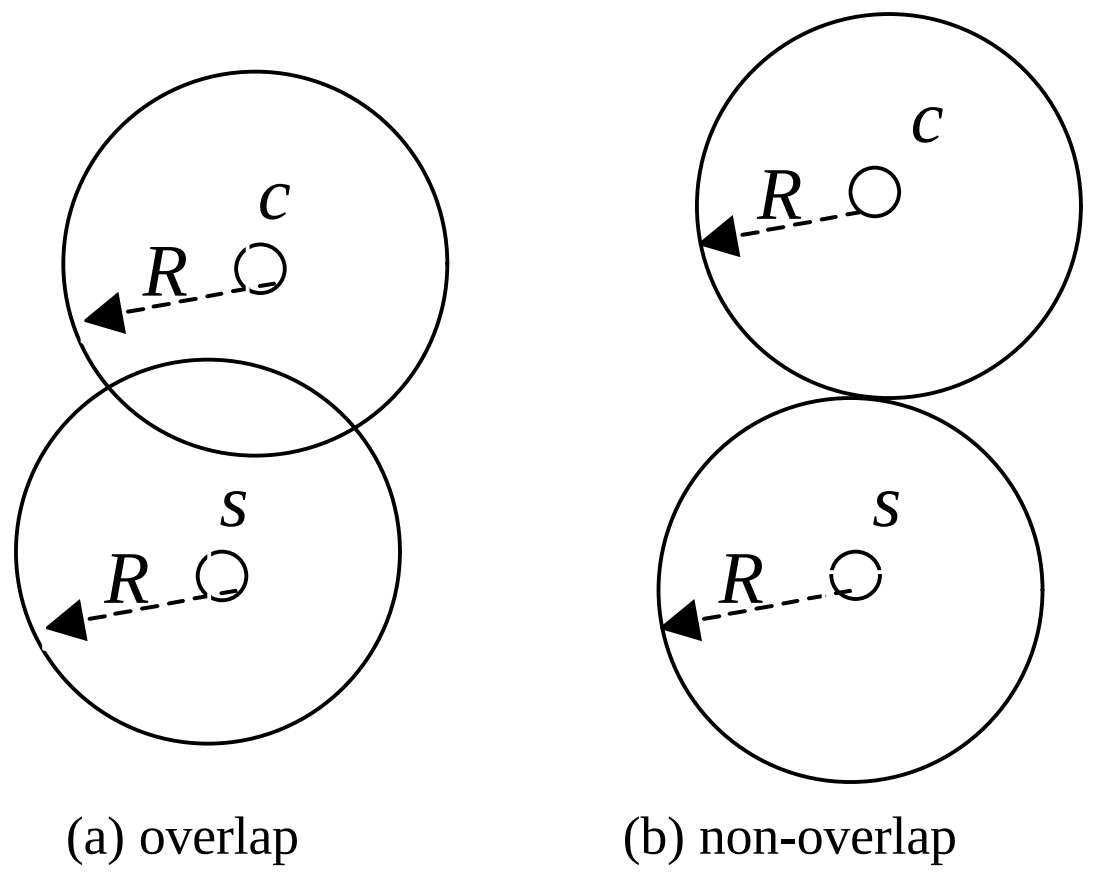

Fig. 1 Overlap in (a) due to the use of *P*

In this figure although the distance between the source and contact is greater than *R* hops, there is still an overlap between the two neighborhoods. Such a situation will arise whenever a node within *R* hops from the edge node becomes the contact. To prevent this situation, equation (*1*) is modified to:

$$P = (d - 2R)/(r - 2R) \quad \text{--(2)}$$

In this equation $P=0$ when $d=2R$ and $P=1$ when $d=r$, i.e., the contact are chosen only between *2R* and *r* hops from the source.

Fig. 2 explains the contact selection procedure with an example. In the above figure *R*=3 and *r*=6. Nodes *a, b, c, d* are the edge nodes for source node *s*. Node *s* sends a Contact Selection Query (*CSQ*) through *a*. Node *a* randomly chooses one of its neighbors *e* and forwards the query to that node. Node *e* calculates the probability *P* according to equation *(1)*. If the probability of being the contact failed at *e*, it forwards the query to one of its neighbors, *f* (chosen randomly). Node *f* again forwards the query to *g*. As *g* is at *r* hops from *s*, the probability *P* at *g* is 1. However, *g* still cannot become a contact for *s* as there already exists another contact *h* (which was selected through a previous query through another edge node *d*) in the neighborhood of *g*. So *g* returns the query to *f* (*backtracking*). Node *f* then forwards the query to another neighbor.

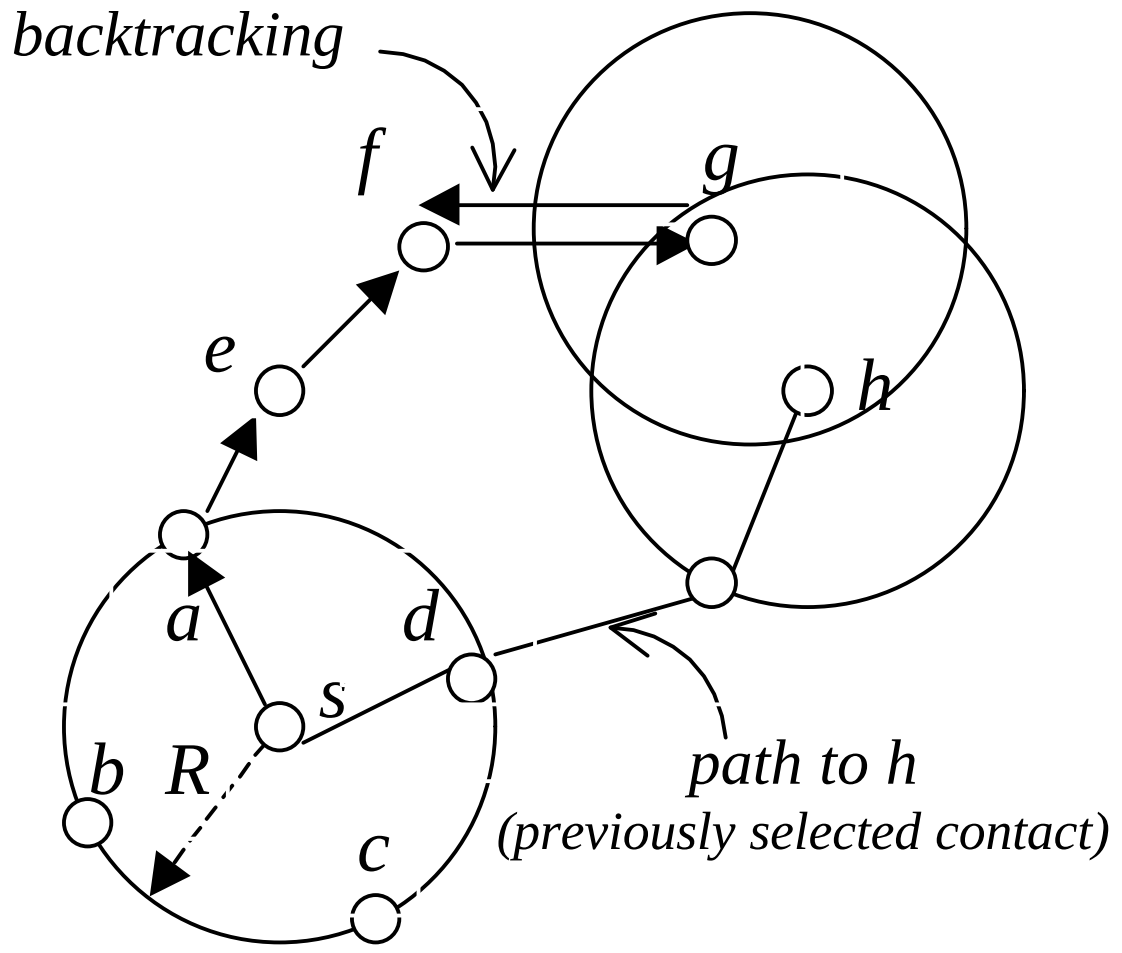

Fig. 2 Selecting contacts in the *P* method

*(b) Edge Method (EM):*

Even with equation (*2*) the probabilistic method can result in a situation where there is some overlap between the contact and the source neighborhoods. This is possible due to the fact that the nodes do not have any sense of direction. Therefore, it is possible that a contact may be selected at a location where the *CSQ* has traveled more than *2R* hops in one direction, but the contact may in fact be closer than *2R* hops from the source. More seriously, the probabilistic method for contact selection can be expensive in terms of the amount of traffic generated by the *CSQ*. This is due to the extra traffic generated due to backtracking, and lost opportunities when the probability fails, even when there is no overlap. To reduce the possibility of such a situation, probability equations (*1*) and (*2*) are eliminated. The probability equations were formulated to have a higher possibility of choosing the contact that lies either between *R* and *r* hops (equation *1*) or between *2R* and *r* hops (equation *2*). To maintain this non-overlapping property without the probability equations, the contact selection procedure is modified as follows.

The list of all edge nodes (*Edge_List*) of the source is added to the *CSQ*. Also, the query and source IDs are included to prevent looping. On receiving a *CSQ*, apart from checking for overlap with the source neighborhood and the neighborhoods of all the already selected contacts (*Contact_List*), the receiving node also checks for overlap with the neighborhoods of any of the nodes on the *Edge_List* as well. Since any node that lies at a distance of less than *R* hops from the edge will have an overlapping neighborhood with the source's neighborhood, checking for non-overlap with the edges ensures that a contact is chosen between *2R* and *r* hops only. This eliminates the possibility of an overlap due to the lack of direction. Fig. 3 and Fig. 4 show a comparison of the probabilistic and edge methods. As can be seen from Fig. 3 the reachability saturates in both *PM* and *EM*. However the saturation occurs much earlier in the case of probabilistic method. Also as compared to *EM*, the reachability achieved is less for *PM*, for the same values of *NoC*. Fig. 4 shows the backtracking overhead for *PM* and *EM*. Due to the reasons explained earlier, the overhead is significantly reduced for *EM*.

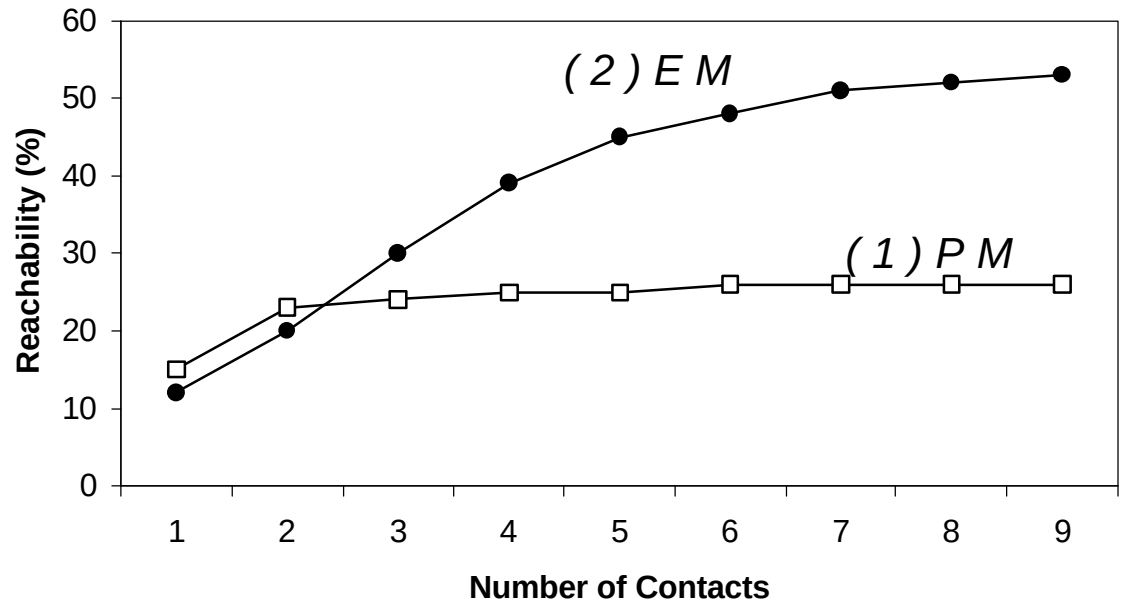

Fig. 3. Reachability for (1) *PM* and (2) *EM*

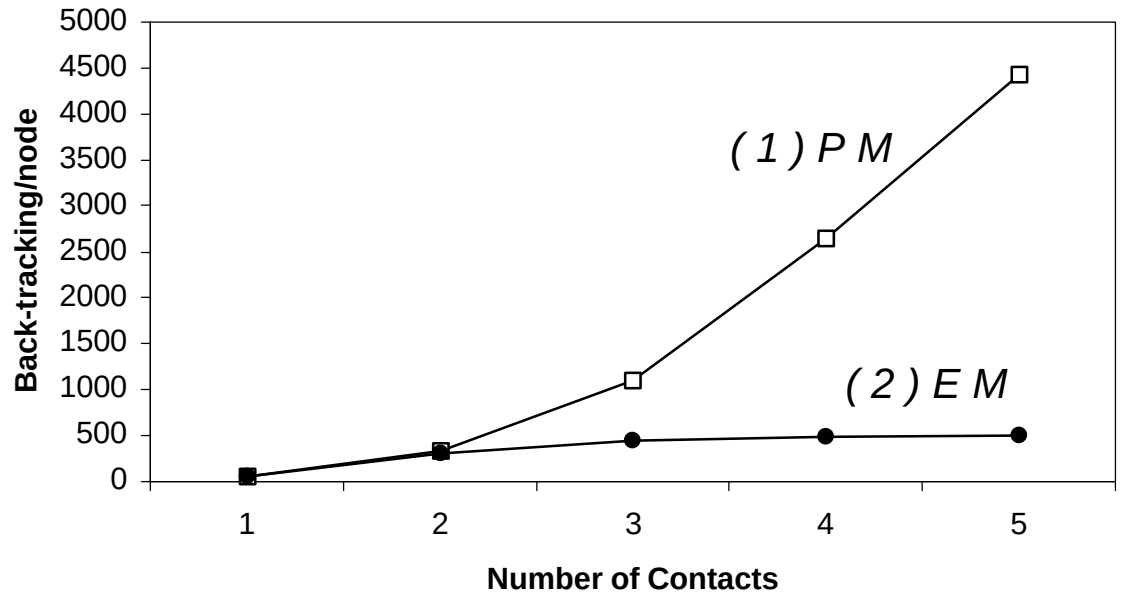

Fig. 4. Overhead for (1) *PM* and (2) *EM*
(Shown for 500 nodes, 710mx710m, tx range=50m, *R=3, r=20, D=1*)

### *3. Contact Maintenance Procedure*

Node mobility may cause the path to a contact to change. Therefore a node needs to keep track of its contacts and their paths. This is done using periodic polling of the contacts as follows.

*(1)* Each node periodically sends *validation* messages to each of its contacts. These validation messages contain the source path to the contact.

*(2)* Each node on the path that receives the validation message checks if the next hop in the source path is a of directly-connected neighbors. If so, it forwards the validation message to the next hop node. If the next hop is missing, the node tries to salvage the path using *local recovery*, discussed later in this subsection.

*(3)* If a path cannot be salvaged using local recovery, the contact is considered to be lost.

*(4)* If the path to a contact is validated but the number of hops to the contact does not lie between *2R* and *r*, the contact is considered to be lost.

*(5)* After validating all the contacts, if the number of contacts left is less than the specified *NoC*, new contact selection is initiated as described earlier.

The *local recovery* mechanism is performed as follows. Assuming reasonable values of node velocities and validation frequency, there is a high probability that if a node has moved out of a contact path, it is still within the neighborhood of the previous hop in the path. Even in the case when a node is completely lost (because it has moved out of the neighborhood of the previous hop), some other node further down the path might have moved into the neighborhood of the previous node. Local recovery takes advantage of these cases to recover from changes in the path, without having to initiate new searches from the source. Thus local recovery provides an efficient mechanism for validating contacts and recovering from changes in the contact paths. If the next hop in a source path is missing, the node that received the validation message looks for the next hop in its neighborhood routing table. If the next hop is in the neighborhood, the source path is updated and the validation message is forwarded to the next hop. If the lookup for the next hop fails, lookup is done for the subsequent nodes in the source path.

### *4. Query Mechanism*

When a source node, ***s***, needs to reach a destination or target resource, *T*, it first checks its neighborhood routing table to see if *T* exists in its own neighborhood. If *T* is not found in the neighborhood, *s* sends a Destination Search Query (*DSQ*) to its contacts. The *DSQ* contains the following information: (1) depth of search (*D*), and (2) target resource ID (*T*). Upon receiving a *DSQ*, each contact checks the value of *D*. If *D* is equal to 1, the contact performs a lookup for *T* in its own neighborhood. If *T* exists, then the path to *T* is returned to ***s***, and the query is considered successful. Otherwise, if *D>1*, the contact receiving the *DSQ* decrements *D* by 1 and forwards the *DSQ* to each of its contact, one at a time. In this way the *DSQ* travels through multiple levels of contacts until *D* reduces to 1.

The source ***s***, first sends a *DSQ* with *D=1* to its contacts, one at a time. So only the first level contacts are queried with this *DSQ*. After querying all its contacts if the source does not receive a path to the target within a specified time, it creates a new *DSQ* with *D=2* and sends it again to its contacts, one at a time. Each contact observes that *D=2* and recognizes that this query is not meant for itself. So it reduces the value of *D* in the *DSQ* by 1 and forwards it to its contacts one at a time. These contacts serve as second level contacts for the source. A second level contact on receiving the *DSQ* observes that *D=1* and it does a lookup for the target *T* in its own neighborhood and returns the path to *T*, if found. In this way the value of *D* is used to query multiple levels of contacts in a manner similar to the expanding ring search. However, querying in *CARD* is much more efficient than the expanding ring search as the queries are not flooded at with different TTLs but are directed to indiviual nodes (the contacts). Contacts leverage knowledge of their neighborhood topology (gained through the proactive scheme operating within the neighborhood) to provide an efficient querying mechanism.

## IV. EVALUATION AND ANAYLSIS

In this section we present detailed simulation based evaluation and analysis of our architecture. NS-2 [14] along with our *CARD* extensions and other utilities were used to generate various scenarios of ad hoc networks. Mobility model for these simulations was random way-point model. Our simulations so far did not consider MAC-layer issues.

First we try to understand the effect of various parameters such as the neighborhood radius (*R*), the maximum contact distance (*r*), the number of contacts (*NoC*), the depth of search (*D*) and the network size (*N*) on (*A*) the reachability and (*B*) overhead. Reachability here is defined as the percentage of nodes that are reachable from a source node. For overhead we consider the number of control messages. We consider overhead due to contact selection and contact maintenance. Having developed a deeper understanding of the various parameter in our architecture, we then compare it with other schemes such as flooding and bordercasting in terms of querying overhead and query success rate.

Table 1 shows the scenarios used in our simulations. These scenarios vary in number of nodes, network size, and propagation range. The variation is considered to capture the effect of these factors on *CARD*. As shown in Fig. 3 and Fig. 4, the edge method outperforms the probabilistic method. (We obtained similar results for other scenarios.) Therefore, we present only the results for the edge method.

TABLE 1 Description of various scenarios used for simulating *CARD*

| No. | Nodes | Area | Tx Range | No. of Links | Node Degree | Network Diameter | Av. Hops |
|---|---|---|---|---|---|---|---|
| 1 | 250 | 500*500 | 50 | 837 | 6.75 | 23 | 9.378 |
| 2 | 250 | 710*710 | 50 | 632 | 5.223 | 25 | 9.614 |
| 3 | 250 | 1000*1000 | 50 | 284 | 2.57 | 13 | 3.76 |
| 4 | 500 | 710*710 | 30 | 702 | 4.32 | 20 | 5.8744 |
| 5 | 500 | 710*710 | 50 | 1854 | 7.416 | 29 | 11.641 |
| 6 | 500 | 710*710 | 70 | 3564 | 14.184 | 17 | 7.06 |
| 7 | 1000 | 710*710 | 50 | 8019 | 16.038 | 24 | 8.75 |
| 8 | 1000 | 1000*1000 | 50 | 4062 | 8.156 | 37 | 14.33 |

### *A. Reachability Analysis*

Reachability Analysis was conducted to understand how contacts help in increasing the view of the network. Here we present results for a topology of 500 nodes spread over area of 710m by 710m. The details can be seen from Table 1, scenario number 5. Similar results were observed for other scenarios.

#### ***1. Varying Neighborhood Size (R)***

Fig. 5 shows the effect of increasing the neighborhood size (*R*) on reachability. As the value of *R* is increased, the reachability distribution shifts towards the right side i.e., more number of nodes achieve higher percentage of reachability. This increase in reachability with the increase in *R* is due to increase in the number of nodes within the neighborhood. However as the value *2R* approaches the maximum contact distance (*r*), the region available for contact selection (between *2R* and *r*) reduces. This results in less number of contacts being chosen. In the Fig 5, when *R=7*, contacts can only be selected between *2R=14* and *r=16* hops from the source. This small region for contact selection significantly reduces the number of contact and hence the reachability distribution shifts back to the left side. At this point most of the reachability is due to the neighborhood of the source only.

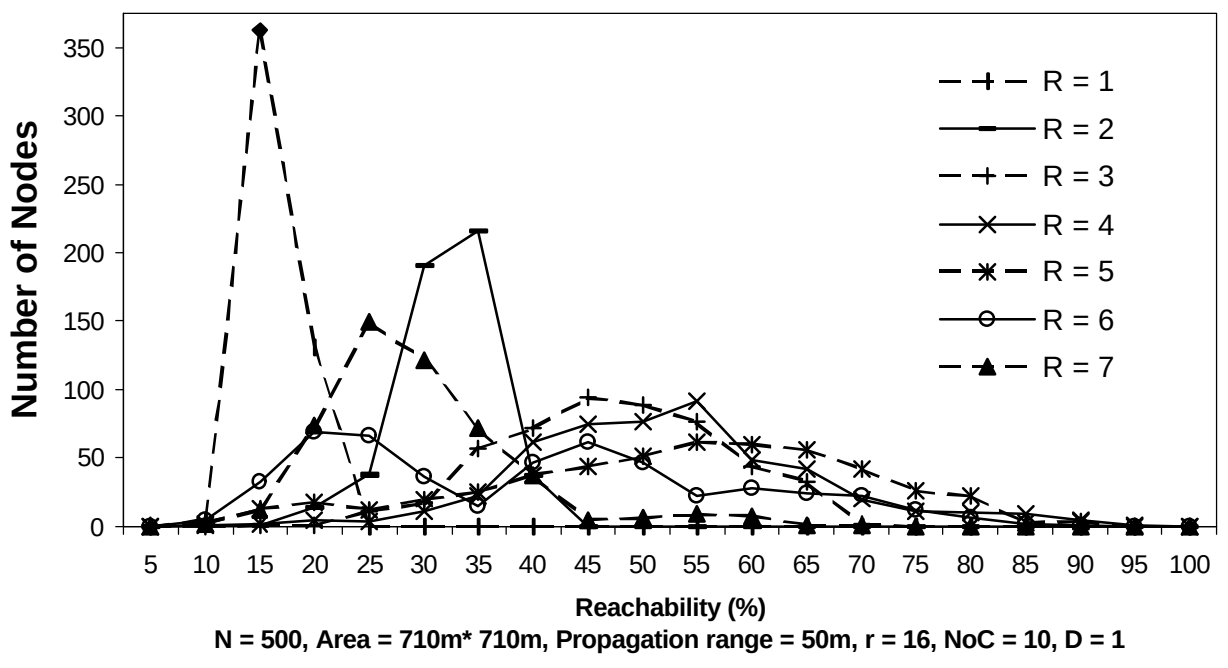

Fig. 5 Effect of Neighborhood Radius (*R*) on Reachability

***2. Varying Maximum Contact Distance (r)***

Fig. 6 shows the effect of increasing maximum contact distance on reachability. Since contacts are selected between *2R* and *r* hops from the source, higher values of *r* provide a wider region for contact selection. The mechanisms for contact selection described earlier prevent selection of contacts that have overlapping neighborhoods. This implies, as *r* increases a larger number of contacts can be selected before their neighborhoods start to overlap. Therefore the reachability increases with increase in *r*. Larger values of *r* also mean that the average contact path length would increase (as more and more contacts are chosen at larger distances from the source). However, once the neighborhoods of the contact and the source are completely non-overlapping, in a uniformly distributed network a far away contact is as good as a near by contact with respect to increase in reachability. Therefore as can be seen from the figure, for *r > (2R +8)* there is not a very significant increase in reachability.

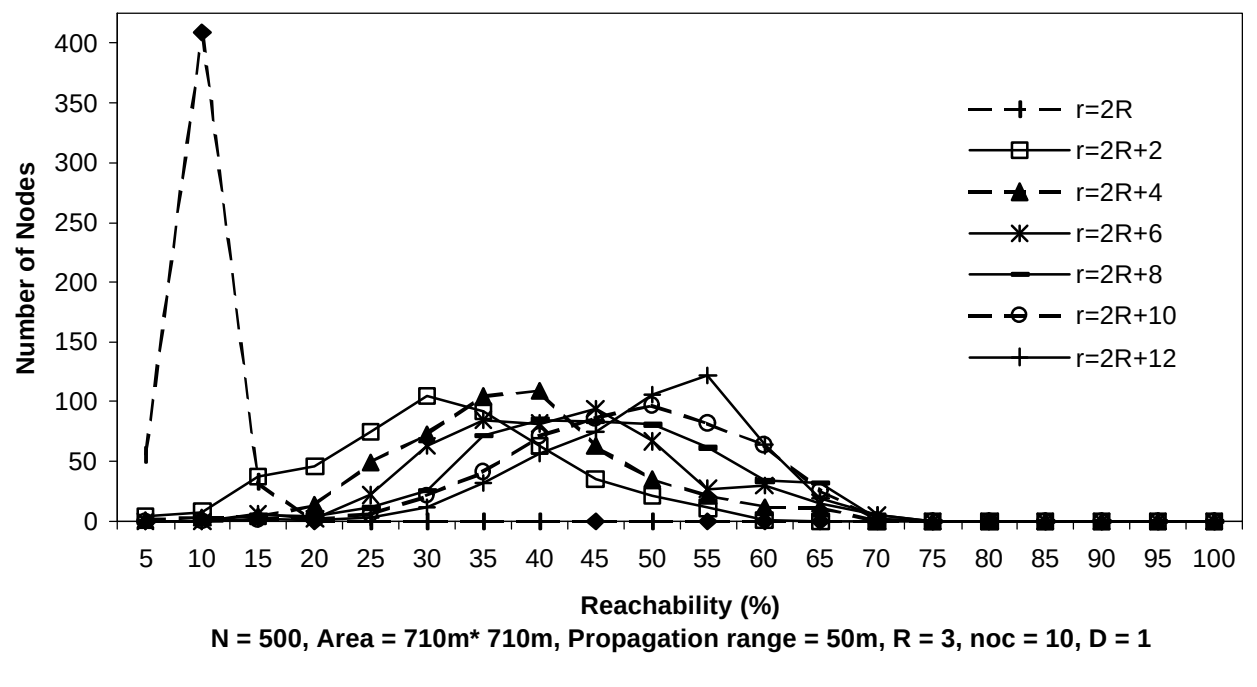

Fig. 6 Effect of Maximum Contact Distance (*r*) on reachability

***3. Varying Number Of Contacts (NoC)***

As described earlier, *NoC* specifies the value of the maximum number of contacts to be selected for each source node. The actual number of contacts chosen is usually less than this value. This is due to the fact that for a particular value of *R* and *r*, there is only a limited region available for choosing contacts. Once this region has been covered by neighborhoods of the chosen contacts, choosing more contacts in the same region is not possible as their neighborhoods would overlap with the neighborhoods of the already chosen contacts. Therefore contact selection mechanism prevents selection of more contacts. This characteristic can be seen in the Fig. 7, in which the reachability initially increases sharply as more and more contacts are chosen. However, the increase in reachability saturates beyond *NoC=6* as the actual number of contacts chosen saturates due to the effect of overlapping neighborhoods described above.

***4. Varying Depth Of Search (D)***

*D* specifies the levels of contacts that are queried in a breadth first manner. When *D=1*, a source node looking for a particular destination beyond its neighborhood, queries its first level contacts only. When *D=2*, if none of the first level contacts contain the destination in its neighborhood, second level contacts (contacts of the first level contacts) are queried through the first level contacts. As can be seen from the Fig 8, reachability increases sharply as the depth of search *D* is increased. Hence, depth of search helps in making a tree-like structure of contacts, which is helpful in making *CARD* scalable.

***5. Varying Network Size***

Fig. 9 shows a variation of reachability distribution for three different network sizes, *N*. The area of the three networks has been chosen so that the node density is almost same across the three networks. Fig. 9 shows that for any given network (specified by the values of *N* and the area), the values of *R* and *r* can be configured to provide a desirable reachability distribution in which most of the nodes have a high value of the percentage reachability.

*B. Overhead Analysis*

Overhead analysis is done in terms of number of control messages required for contact selection and maintenance. The total overhead considered includes:

1. Contact selection overhead: This is the amount of *CSQ* traffic generated for selecting new contacts. This includes overhead due to Backtracking as described earlier.
2. Contact maintenance overhead: This is the traffic generated by the contact path validation messages. *Local recovery*, as described earlier, helps in reducing this part of the total overhead.

Results are shown for scenario number 5 in Table 1. Similar results were obtained for other scenarios.

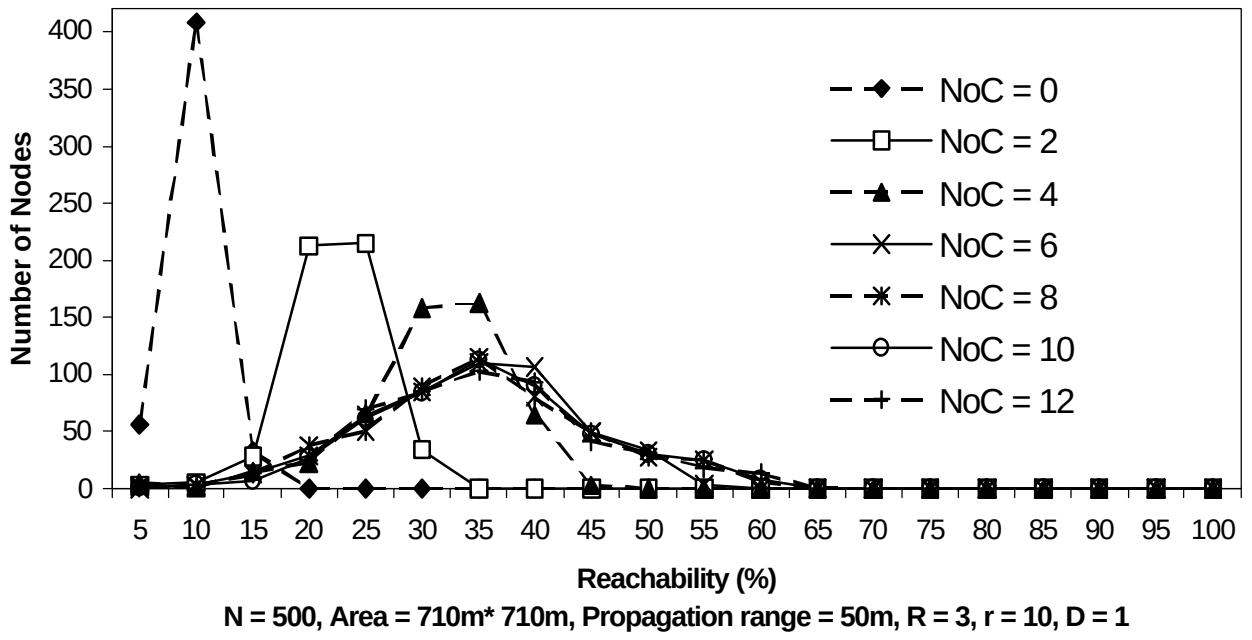

Fig. 7 Effect of Number of Contacts (*NoC*) on Reachability

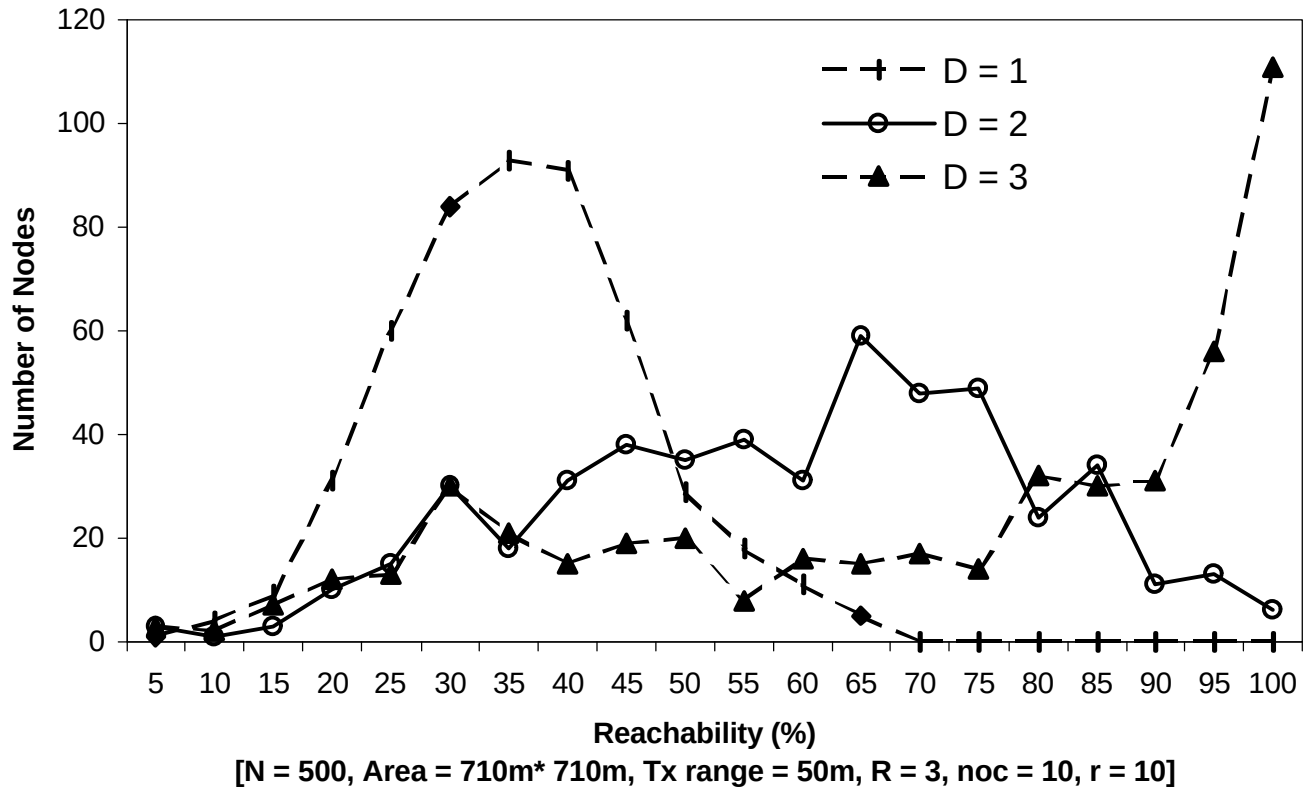

Fig 8. Effect of Depth of Search (*D*) on Reachability

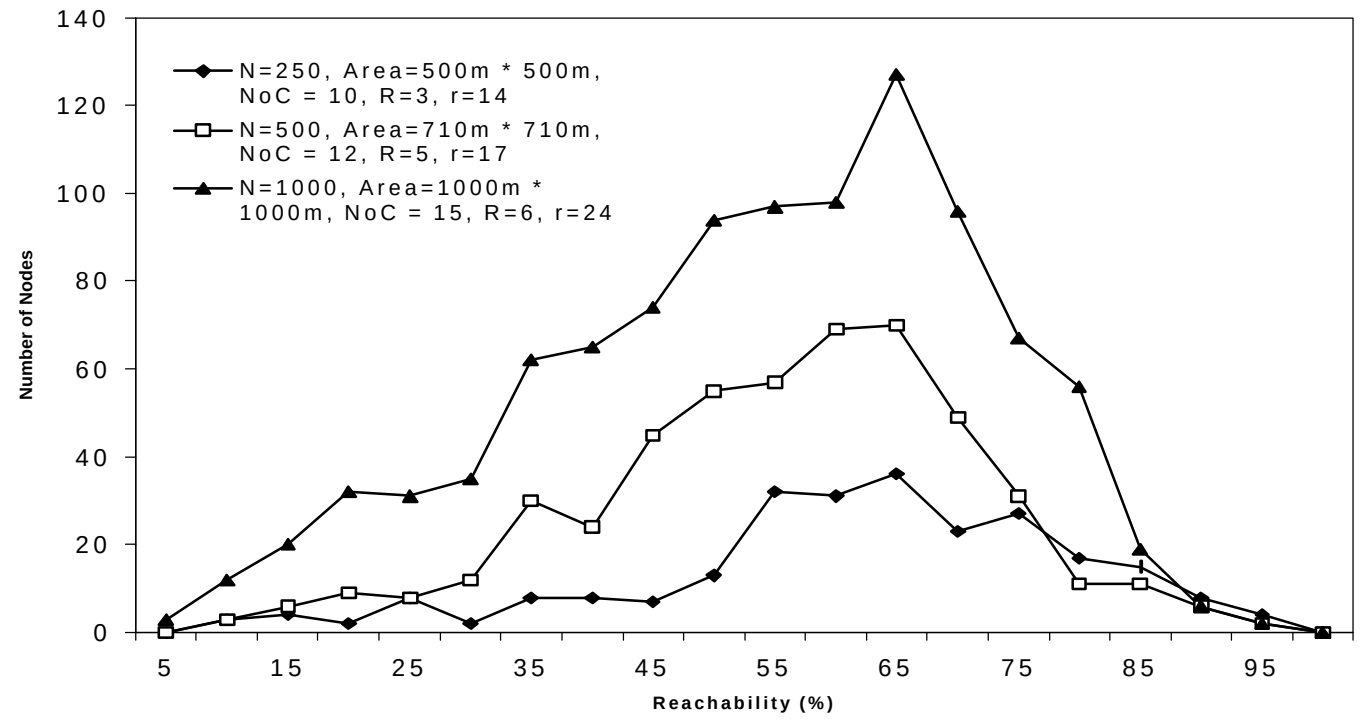

Fig. 9 Reachability for different network sizes.

### *1. Varying Number Of Contacts (NoC)*

As shown in Fig. 10, as the number of contacts increases the maintenance overhead increases sharply as more and more nodes have to be validated.

### *2. Varying Maximum Contact Distance (r)*

As *r* increases the number of contacts increases as shown in Fig. 11, the increase in the number of contacts is due to the availability of a wider area for choosing contacts. Moreover, with higher values of *r*, contacts may lie at greater distances from the source. That is, the contact path length is expected to be higher for larger values of *r*. This suggests that the maintenance overhead should increase with increase in *r*. However as shown in Fig. 11, the overhead actually decreases with increase in *r*. Fig.12 explains this decrease in maintenance overhead. Fig.12 shows that as the value of *r* increases the backtracking overhead decreases significantly. Recall that backtracking occurs when a node receiving a *CSQ* cannot become a contact due to overlap with already existing contacts. As *r* increases, the possibility of this overlap decreases due to availability of a wider area for contact selection. This decrease in back-tracking overhead is significantly more than the increase in overhead due to increased number of contacts and contact path length. Therefore, the total maintenance overhead decreases.

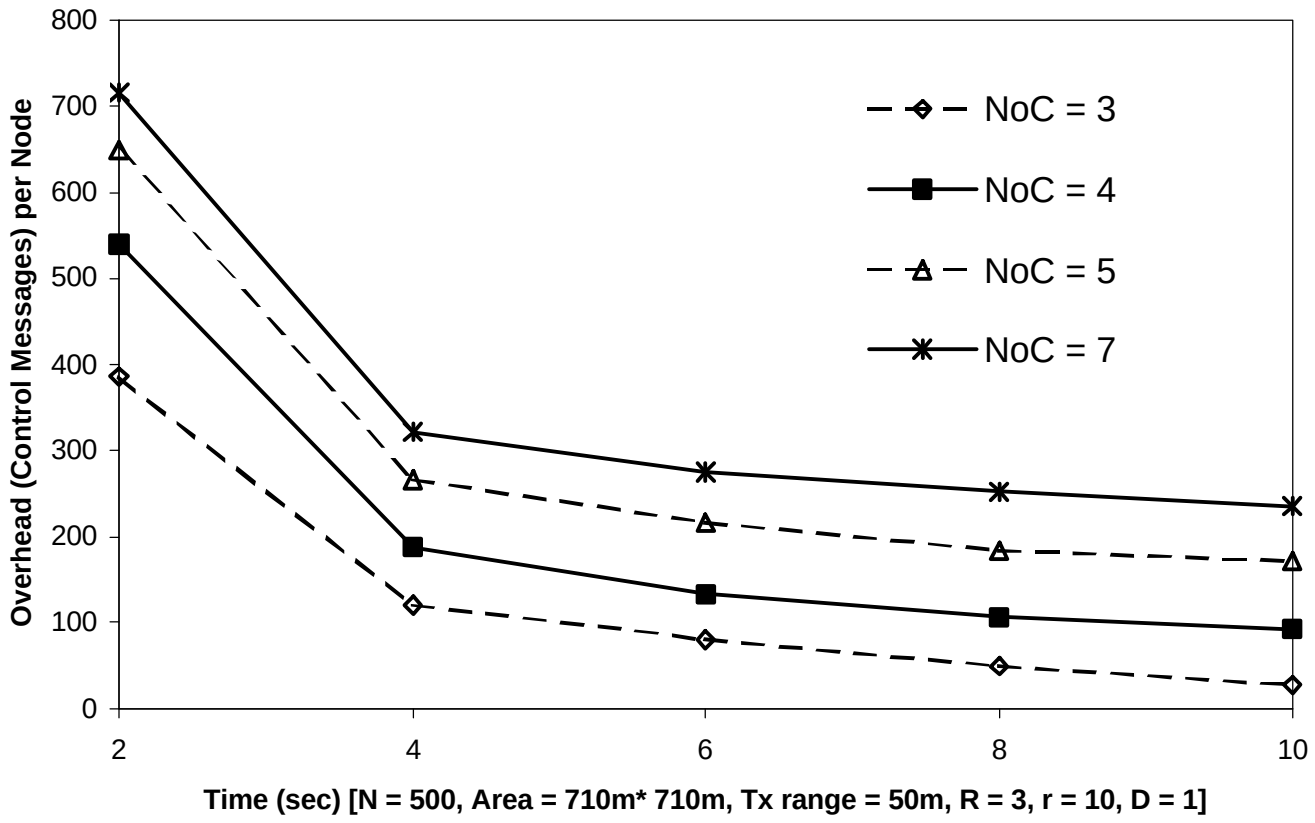

Fig. 10 Effect of Number of Contacts (NoC) on Overhead

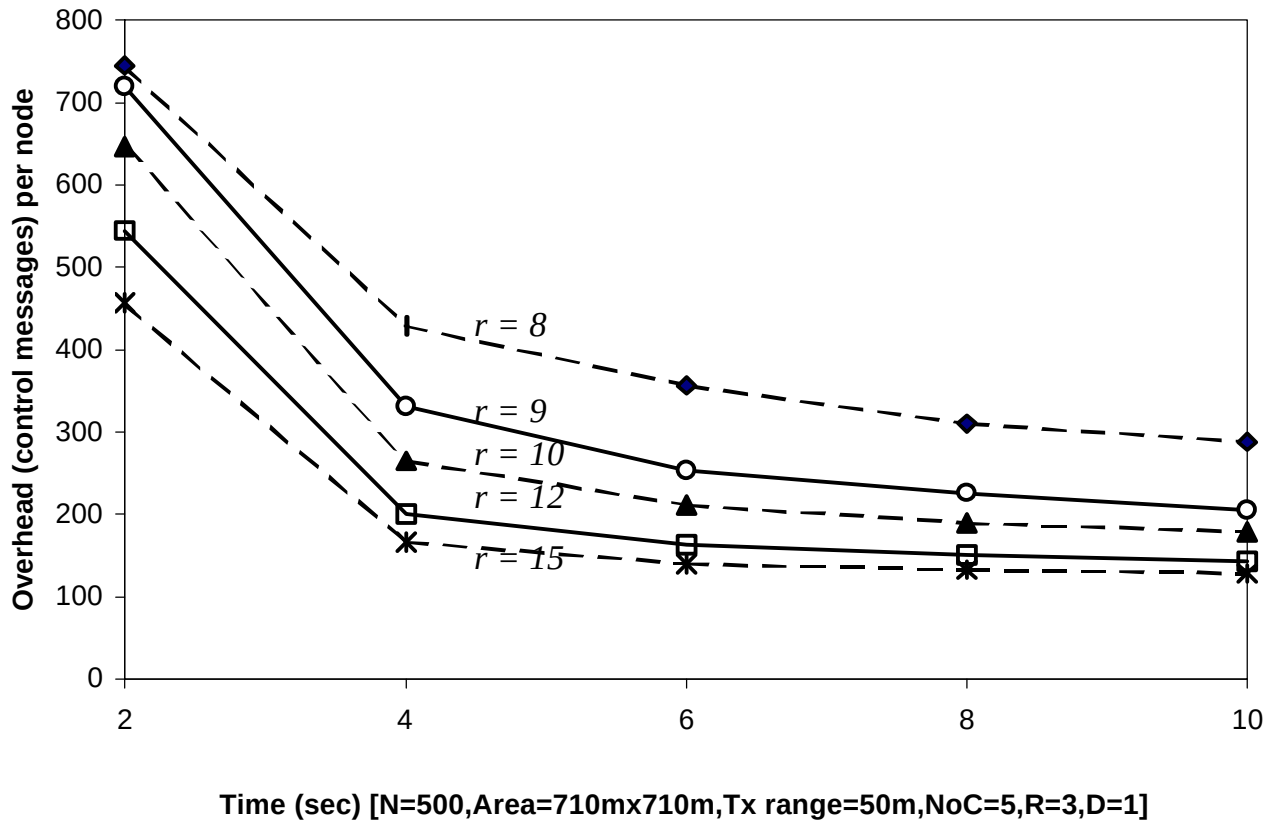

Fig. 11 Effect of Maximum Contact Distance (*r*) on Total Overhead

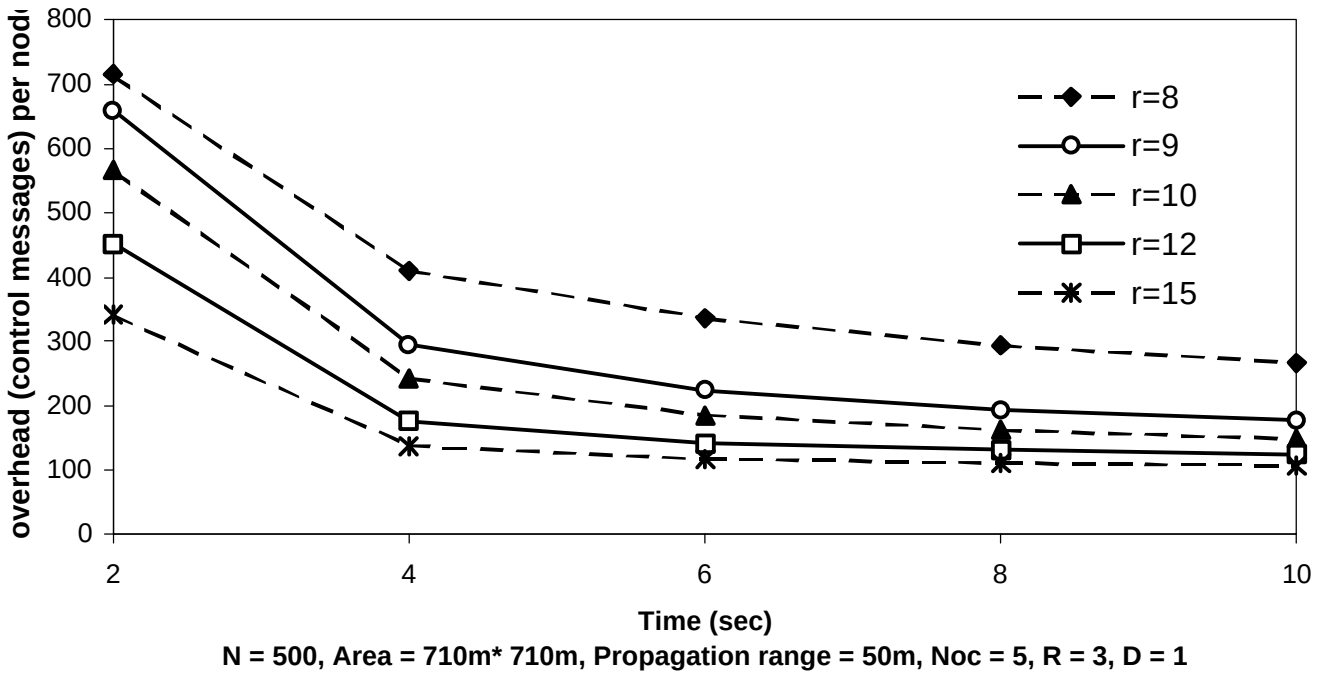

Fig. 12 Effect of maximum contact distance (*r*) on backtracking overhead

### *3. Maintainance Overhead Over Time*

Fig. 13 shows the maintenance overhead per node over a 20sec period. The maintenance overhead decreases steadily with time. However, the number of contacts increases slightly. This suggests that the source nodes find more and more stable contacts. Stable contacts may be defined as those nodes that have low velocity relative to the source node. Therefore, a node moving in the same direction as source node with similar velocity could prove to be a stable contact. Hence, *CARD* leads to source nodes finding more such nodes in the vicinity[1].

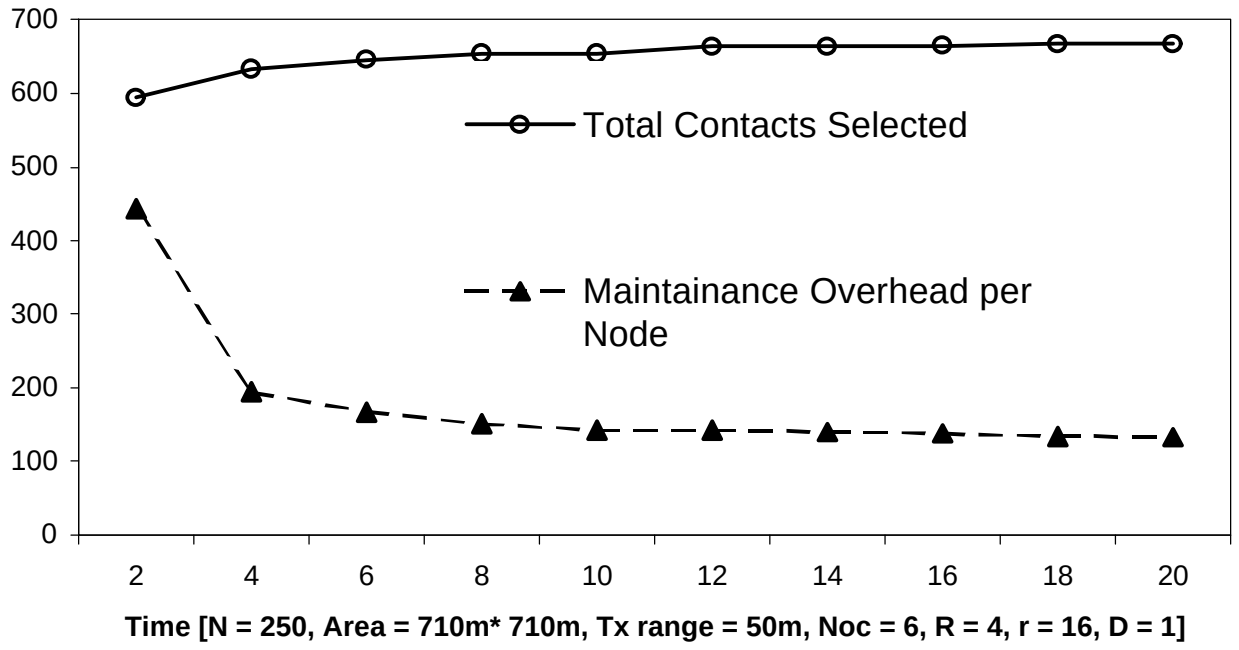

Fig. 13 Variation of overhead with time

### *C. Trade-off Between Overhead and Reachability*

As shown in earlier results, higher reachability maybe obtained by increasing the number of contacts. However, increase in the number of contacts also increases the contact selection and maintenance overhead. Fig. 14 shows this trade-off between reachability and overhead. The figure shows that there exists a desirable region in which good reachability (greater than or equal to 50 %) can be achieved at a reasonable amount of selection and maintenance overhead.

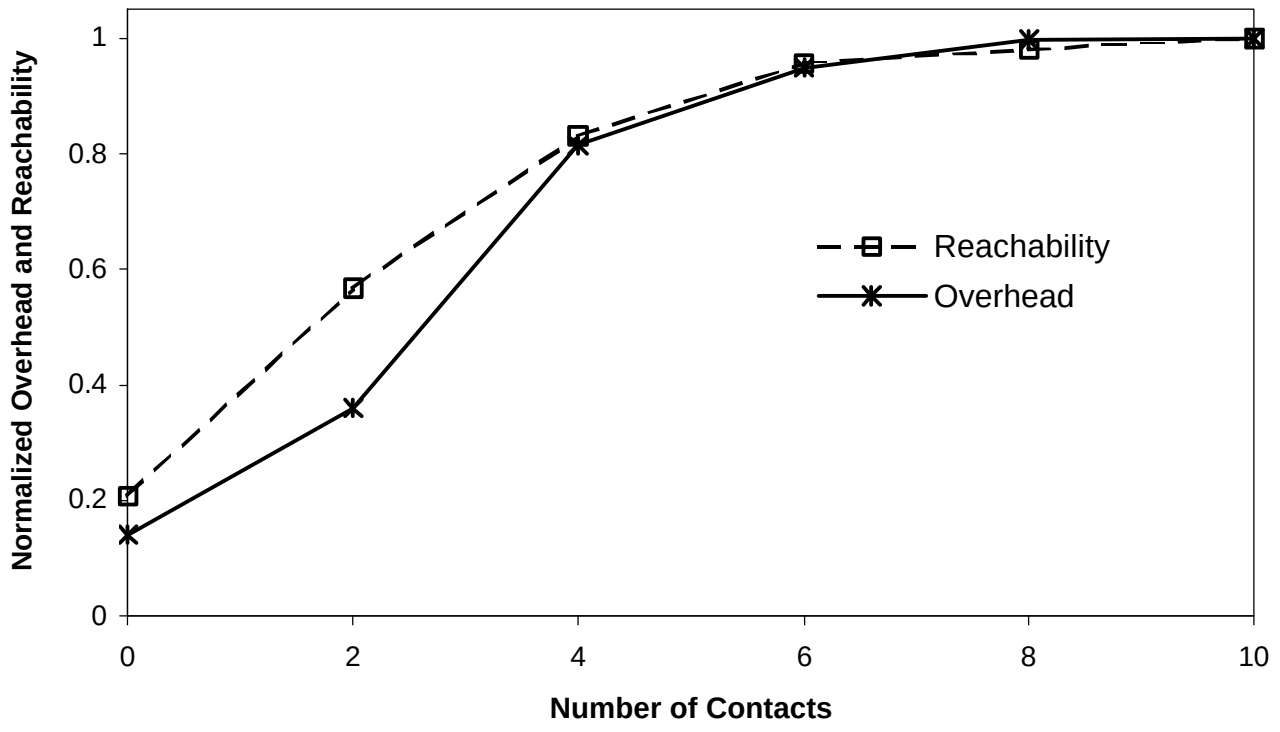

Fig. 14 Trade-off between reachability vs. contact selection and maintenance overhead

### *D. Comparison with Other Approaches*

Here, we present comparison of *CARD* with flooding and bordercasting [8]. Bordercasting was implemented with query detection (*QD1* and *QD2* ) as described in [8].

Fig. 15 shows the average traffic generated per node for querying 50 randomly selected destinations from 50 random sources. *CARD* generated far less querying traffic than the others. Fig. 15 also shows contact selection and maintenance overhead in *CARD*. As can be seen from the figure, *CARD* is still more efficient than the other two approaches. Whereas flooding and bordercasting result in 100% success in queries, *CARD* showed a 95% success rate with *D=3*. *CARD's* success rate can be increased by increasing *D*.

[1] This may also be due to random way-point (RWP) model. We plan to investigate this problem further and expect that different mobility models may have different effects on performance of *CARD*.

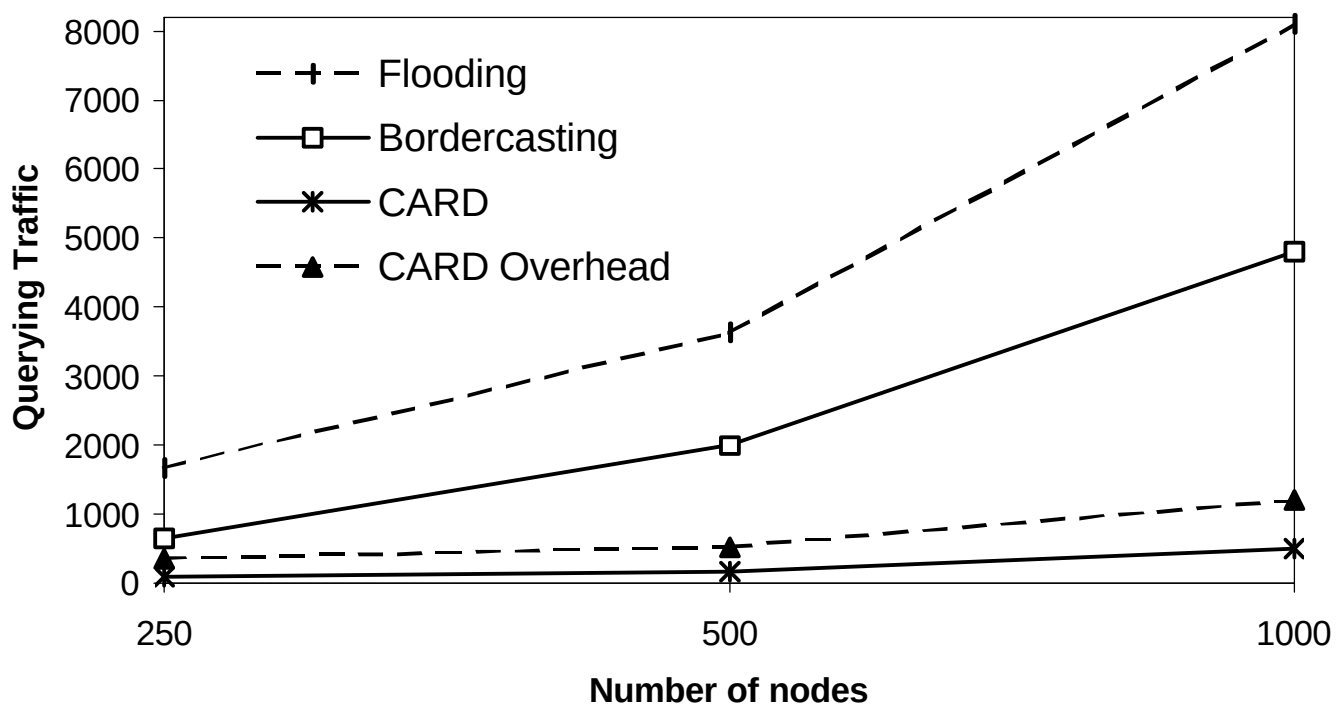

Fig. 15 Comaprison of CARD with other scheme

## V. CONCLUSIONS

In this paper we presented a novel architecture for resource discovery in large-scale ad hoc and sensor networks. Salient features of our architecture include its ability to operate without requiring any location information or any complex coordination. In our architecture, each node knows routes to nodes within its neighborhood. Based on small world concepts, we have introduced the notion of *contacts* to serve as short cuts that extend the view of the network per node and increase reachability beyond the neighborhood. Two protocols for contact selection were introduced and evaluated: (a) probabilistic method and (b) edge method. We evaluated the protocols for reachability and contacts selection and maintenance overhead under mobility conditions. The latter approach was found to experience less overhead during selection due to reduced backtracking, and was thoroughly analyzed over the various dimensions of the parameter space (including *R*, *r*, *D*, *NoC*, and network size). We further compared our approach to other resource discovery techniques; flooding and bordercasting. The overall overhead experienced by the contact-based architecture was found to be significantly lower than the rest. These results show a lot of promise for the contact-based approach and we are encouraged to further investigate this direction. We plan to further evaluate our protocols under various scenarios of mobility patterns and resource distributions in the network. We shall also pursue other heuristics for contact selection mechanisms.